\documentclass[aip]{revtex4-1}
\usepackage{graphicx, amsmath, amssymb}

\begin{document}
 
\title{Complex dielectric properties of organic-rich mudrocks as functions of maturity}       

\author{Rezwanur Rahman}

\affiliation{Department of Physics, Colorado School of Mines, Golden, CO 80401-1887, USA}
\affiliation{Department of Petroleum Engineering, Colorado School of Mines, Golden, CO 80401-1887, USA}
\affiliation{OCLASSH, Department of Petroleum Engineering, Colorado School of Mines, Golden, CO 80401-1887, USA}
\author{John A. Scales}

\affiliation{Department of Physics, Colorado School of Mines, Golden, CO 80401-1887, USA}
\author{Manika Prasad}

\affiliation{Department of Petroleum Engineering, Colorado School of Mines, Golden, CO 80401-1887, USA}
\affiliation{OCLASSH, Department of Petroleum Engineering, Colorado School of Mines, Golden, CO 80401-1887, USA}

\begin{abstract}

     Complex dielectric variations can address neatly the maturity of organic-rich mudrocks. We, therefore, apply an open hemispherical cavity resonator
    to measure complex dielectric permitivitties of five thin sections of oil (bakken) shales (with different maturity) of $\sim$ 30 $\mu$m thickness
    on glass-substrates around 2.15 mm thick in 100 - 165 GHz. The real part of complex dielectric permittivity ($\epsilon_{re}$) are constant but show significant
    differences in magnitude based on maturity: (1) lowest, $\sim$ 1.9, for immature or early maturation stage, (2) higher than the immature one, $\sim$ 3.0 (also bundled up together),
    for three oil-matured stages, and (3) highest, $\sim$ 4.9, for late or overmatured stage. The conductivity ($\sigma_{re}$) from imaginary part of the 
    complex dielectric constant ($\epsilon_{im}$) emphasizes two important features of conductivity of oil shales: (1) frequency dispersion, 
    and (2) maturity-dependence. We obtain increases of $\sigma_{re}$ with maturity and frequency. 
    

 \end{abstract}

\maketitle 

\section{Introduction}

 Energy demands have fueled development of so-called shale reservoirs. This development is two-fold: (i) subsurface oil and gas
production from mature (naturally retorted) oragnic matters (OM) and (ii) near-surface oil and gas generation by artificial retorting
of immature OM. By tracking maturity through indirect sensing, we can be more conservative in explorative drilling and fracking.
Similarly, by tracking soil contamination and leakage, during surface retorting or completion of and production from boreholes,
energy from shale can be green (clean and non-toxic). Despite mature technology to produce hydrocarbon (HC) from shales,
their organic maturity can not be measured without physically extracting a sample.

We investigate the maturity dependence of complex dielectric constant, $\tilde{\epsilon}$, of organic-rich rocks.
Our main objective is to develop a reliable relation between $\tilde{\epsilon}$ and maturity
to be used in electromagnetic (EM) surveys of shale reservoirs.
Understanding $\tilde{\epsilon}$ = $\epsilon_{re}$ + i$\epsilon_{im}$ of a rock
is challenging as it comprises the effects of mineralogy, porosity, pore fluids, frequency, geometry,
and electrochemical interactions between fluid and rock components.\citep{AM01} In addition, for organic rich-mudrocks, dielectric loss 
($\epsilon_{im}/\epsilon_{re}$) varies also 
as a function of kerogen content and its maturity.\citep{JJS07}
DC conductivity, related to $\epsilon_{im}$, of source rock can be instrumental to be used as a measure of the presence of hydrocarbons.\citep{PN83}

Mudrocks and other nano-porous reservoir rocks (also called unconventional rocks) consist of some mixture of biogenic sediments
(kerogen and carbonate and siliceous oozes) as well as non-biogenic minerals (clay, quartz,
calcite, and dolomite).\citep{STW80, PA71} Key differences between conventional and unconventional reservoirs
lie in the role of surfaces, surface chemistry, micron-sized grains, nanometer-sized pores, and low, nano 
Darcy-range permeability.\citep{QRP10} For example, there is a 10-fold or larger increase in surface area in unconventional reservoir
rocks as compared to conventional rocks, being not limited to clay minerals.\citep{LCM14}
Detection and quantification of these small pores can pose unique
challenges.\citep{UK13}

Understanding material properties of shales is quite a bit complex as (1)
hydrocarbon generation from organic matter is only one of the maturation processes\citep{JH05, LB94}; (2) inorganic 
minerals also undergo diagenesis and maturation\citep{JSS03}; (3) fluid content and type change
with maturation.\citep{QRP10}

The complex dielectric properties of dry and saturated Green
River oil shale measured in the frequency range, 1 MHz - 1.8 GHz.\citep{JJS07}
presents three important impacts : $\tilde{\epsilon}$ of (i) dry rock
almost constant with frequency (accompanied by very low values of $\epsilon_{im}$) and (ii) saturated rock increase with de-ionized
water and brine, and (iii) temperature-dependence on $\tilde{\epsilon}$ is minimal.
Kerogen-dependence of dielectric constant of Green River shales in millimeter-wavelength ranges is lower in
organic-rich rock than that of an organic-poor one.\citep{JS0689}
On the other hand, scientific investigations of Jordonian oil shales show no correlations between dielectric properties and total 
organic matter contents.\citep{MAl09}
At $\textit{in situ}$ temperature conditions, complex dielectric permittivity (of Green River oil shale) shows
that at low temperatures ($<$200$^{0}$C) and constant oil grade, complex dielectric permittivity is inversely proportional to
frequency and level off at higher ($\sim$ 10$^{12}$ Hz) frequencies.\citep{JAH11}
\citet{JAH11} show that real part of the complex dielectric permittivity increases with temperature; its
imaginary part fluctuates in the higher temperature zones ($>$200$^{0}$C).
Anisotropic complex eletctrical conductivity of oil shales (Bakken) confirms its
dependencies on saturation, salinity, porosity, organic maturity and CEC.\citep{AR13}
Dielectric spectroscopy (for both preserved shales and shale cuttings) at high frequency (GHz)
can be instrumental to obtain water content without any information of water salinity.\citep{Cl06}
The time domain THz spectroscopy (TDS-THz) applied to probe 
vitrinite reflectance ($R_{o}$) of mudstones as functions of absorption coefficient and kerogen contents at different maturities at a single
frequency; they demonstrate that $R_{o}$ increases with temperature (between 300 and 500 K) indicating also the increase
of maturity. \citet{BR13} also show that abosrption coefficient peaks at maximum for $\sim$ 0.8 $R_{o}$ (for crude oil) and at much lower one for 
$\sim$ 1.6 $R_{o}$ (for natural gas) where different maturities are attained by pyrolysis.

At $\textit{in situ}$ conditions, electromagnetic charaterizations of oil shales help in designing EM well logging
tools.\citep{JAH11} The low frequency ($\sim$ Hz) dielectric data can be implemented to calibrate downhole log data.\citep{Cl06}
Dielectric dispersion investigations would allow us to develop a dielectric logging tools for measuring 
constrasts of dielectric permittivity data of constituents of source rocks.\citep{NVS11}

Therefore, dielectric properties as a function maturity of oil shales, is capable of illustrating this correlation
just because $\epsilon_{im}$ or $\sigma_{re}$ can address the loss mechanisms due to porosity and flow of free radicals, and the 
$\epsilon_{re}$ can indicate the kerogen contents.

The oil generating source rocks and humic soil illustrate the interesting contrast in terms of complex dielectric
permittivity. \citet{KT14} measure the dielectric constant and absorption coefficient of humic soil with
GPR (Ground-penetration radar); the result shows that its dielectric permittivity ($\epsilon_{re}$) is high
around 20 and absorption coefficient (at 1.5 GHz) remains low, 7.69 - 9.96 ($\Omega^{-1}$ cm$^{-1}$). 
The dielectric permitivitties of heavy (crude) and light
(crude) oil are found to be $\sim$ 3 and $\sim$ 2.\citep{} Asphalt (soild) shares very close value to light
oil, which is $\sim$ 2.6 at 76 F.
A study on complex dielectric properties of bitumen by \citet{XJ85} showed $\epsilon_{re}$
and $\sigma_{re}$ of bitumen increase with the degree of its metamorphism.
To our knowledge, dielectric permittivity as a function maturity in natural organic source rocks are missing.
In this paper, we present the first complete data of complex dielectric permittivity of mudrocks (Bakken) as a function of maturity, which
is critically related to HI and metamorphism of bitumen.
We investigate the impacts of these phenomena to the conductivity
(inverse of resistivity) of source rocks in the oil window. Organic maturation processes are known to generate
polar and non-polar HC species. Our previous measurements on montmorillonite clay have shown that $\epsilon_{re}$
is higher in a monovalent cation (Na$^{+}$) stabilized-montmorillonite clay than in a bivalent cation (Ca$^{++}$)-stabilized one.\citep{RRC14}
We anticipate similar difference in OM depending on organic molecules polarity and ionic strength.

\section{Materials}
We studied thin sections of low porosity ($\Phi$ $<$ 5$\%$) organic rich-mudrocks from the Bakken formations (North Dakota). 
Details of the samples are described\citep{LV97,LV96}; HI (hydrogen index) decreased from 584 to 161 with increasing depth and is
used as an indicator of organic maturity, with decreasing HI as maturity-increase.

\begin{table*}
\caption{\label{table 1}In $\sim$120 GHz, the real parts of the complex dielectric constants ($\epsilon_{re}$)  and conductivities ($\sigma_{re}$) of five bakken shales}
\begin{ruledtabular}
\begin{tabular}{cccccc}
(Kerogen shale)&Depth&Maturity& Description& \multicolumn{2}{c}{120 GHz}\\
                                             \cline{5-6} 
               &          &    &             &$\epsilon_{re}$  &$\sigma_{re}$  \\
Bakken           &(m)    &Stage&          &                     &($\Omega^{-1}$cm$^{-1}$)\\ \hline 
 1     &2630          &II & Onset HC generation                        &1.9 &0.012                                 \\
 2     &3098          &III & Advanced HC generation                    &3.0 &0.021                                 \\
 3     &3223          &IVa & Main-stage HC generation;primary migration     &3.2 &0.022                        \\ 
 4     &3332          &IVb & Main-stage HC generation;primary migration     &3.4 &0.024                        \\ 
 5     &3428          &V  & Condensate and wet gas                   &4.9 &0.034                                    \\ 

\end{tabular}
\end{ruledtabular}
\end{table*}

\section{Methods}
  
We used an open hemispherical cavity resonator,\citep{RR13a,Dudorov95} to measure
complex dielectric permittivity for samples much smaller or thinner than the wavelength of the probing
signal, in our case, Bakken thin sections. Detail mechanical design of a open hemispherical cavity, its structure and method of measuring
complex dielectric properties are explained in \citet{RR13a}

\section{Results and Discussions}

In the article, we present dielectric permittivity, $\epsilon_{re}$, and electrical conductivity, $\sigma_{re}$, of
five Bakken thin sections with varied maturity, at frequencies 100 and 165 GHz (3.0 to 4.95 cm$^{-1}$) at room temperature.
The $\epsilon_{re}$ data of all these organic-rich mudrocks are constants in frequency (no dispersion) but hit the highest value, $\epsilon_{re} \sim$ 4.9, 
for late mature stage, V, and the lowest one, $\epsilon_{re} \sim$ 1.9, for early maturation stage, II. In advanced hydrocarbon (HC) 
generation stages, III, IVa and IVb, $\epsilon_{re}$ remain almost constant but clustered, range $\sim$ 3.0 - 3.4. Speaking of $\sigma_{re}$, it shows 
the similar patterns as $\epsilon_{re}$ except frequency dispersion; all five stages follow sublinear ($\nu^{p}$, 1$\lesssim p$) trends and only V contains 
a constant part at lower frequencies. The particular $\nu^{p}$-pattern in conductivity has a significance we will discuss later in this section.

In order to correlate these electrical parameters, $\epsilon_{re}$ and $\sigma_{re}$, to maturity, key components should be
(1) the properties of kerogen-bitumen-clay (KBC) complex, (2) its interactions with minerals, and (3) its conversion into polar/non polar 
hydrocarbons. Due to low porosity ($<$ 5$\%$) profiles of our Bakken samples, it is difficult to establish any porosity dependence.
\citet{LV96} describe the type of OM's with their level of metamorphisms during the variuos maturity stages.

Kerogen is a complex consisted of decomposed algae buried in fine-grained sediments under anaerobic conditions.\citep{BR13}
Chemical structure of kerogen varies from sample to sample that makes it almost impossible to characterize. Bitumen is defined
as a converted organic matter (OM) which contain the by-products of kerogen. Therefore, metamorphisms of kerogen-bitumen-clay (KBC) complex
play a major role for identifying maturation processes. However, we use frequently KBC complex as 
there is no solid demarcation line between kerogen and bitumen.\citep{ZR15}

\begin{figure}
\centerline{
    \includegraphics[width=.5\textwidth]{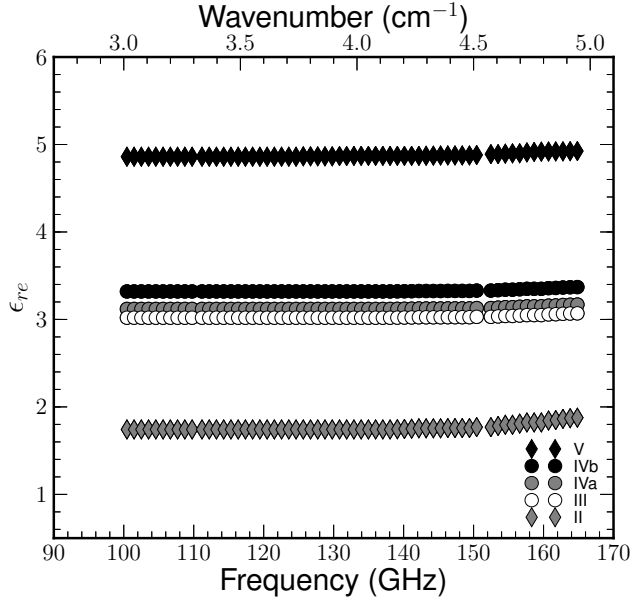}}
  \caption{Real part of dielectric constant ($\epsilon_{re}$) data: Bakken oil shales (thin sections) with different maturity
           between 100 and 165 GHz.}
  \label{reeps_BK}
 \end{figure}

 \begin{figure}
\centerline{
    \includegraphics[width=.5\textwidth]{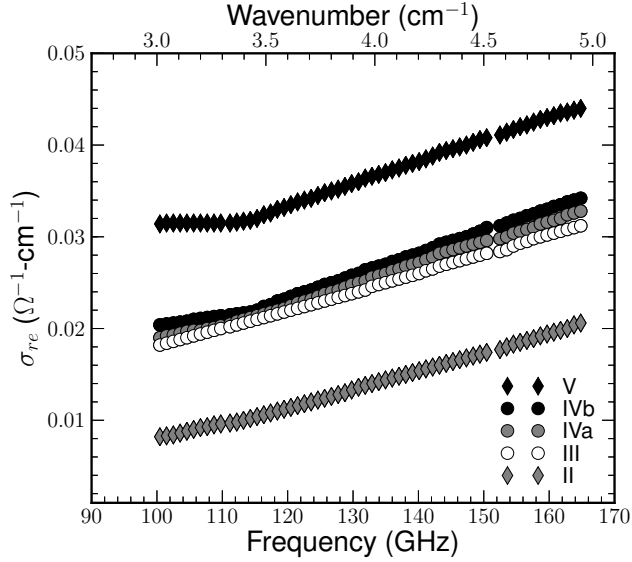}}
  \caption{Experimental conductivity-data: Bakken oil shales (thin sections) with different maturity
           between 100 and 165 GHz.}
  \label{optcond_BK}
 \end{figure}

 \begin{figure}
\centerline{
    \includegraphics[width=.5\textwidth]{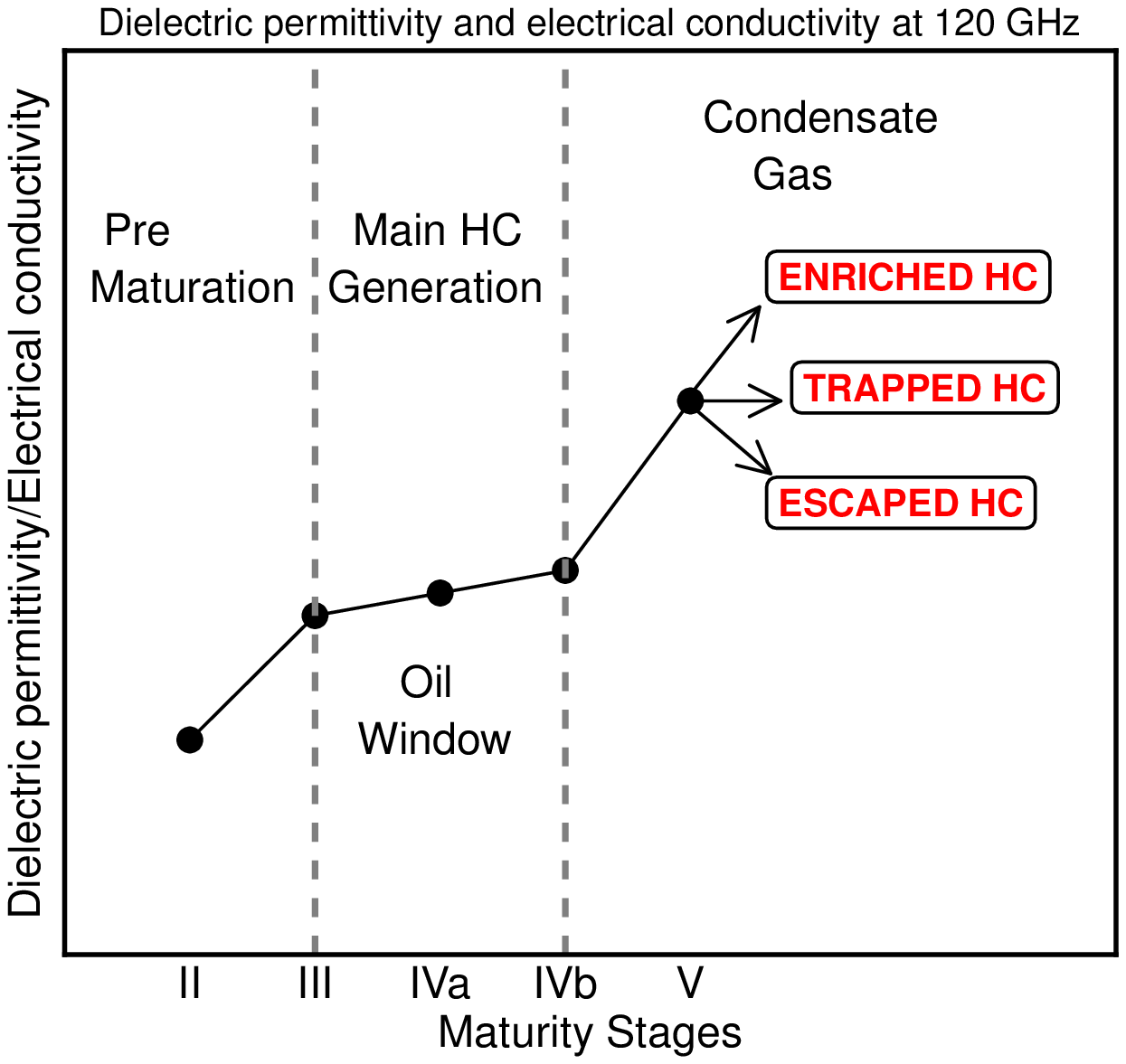}}
  \caption{Schematic diagram of dielectric permittivity and electrical conductivity as functions of maturity (as $\epsilon_{re}$ and $\sigma_{re}$ follow exactly same 
           pattern): Bakken oil shales (thin sections) with different maturity at 120 GHz.}
  \label{maturity_BK}
 \end{figure}

\citet{XJ85} demostrates microwave complex dielectric constant of bitumen shifts from low to high depending on oil transformation phases. 
It\citep{XJ85} shows that bitumen has $\epsilon_{re}$ $\sim$ 2.6 and $\sigma_{re}$ $\sim$ 0.023 ($\Omega^{-1}$cm$^{-1}$) in the oil-generating
areas ; in the oil-pyrolyzed gas areas, these can attain very high-values, $\epsilon_{re}$ $\sim$ 5-15, and $\sigma_{re}$ $\sim$ 0.85 
($\Omega^{-1}$cm$^{-1}$). It is reported that thermal metamorphic degree of bitumen is usually small so that $\epsilon_{re}$ and $\sigma_{re}$ 
are low.

Our results (Table \ref{table 1}) of dielectric permittivity and conductivity for organic-rich mudrocks in the HC generating phases, are consistent with
the data found for the bitumen\citep{XJ85}(in the same oil-phases). The KBC complex contrubites to dielectric permittivity, 
$\epsilon_{re}$ of oil-rich mudrocks; their maturity enhances conductivity, $\sigma_{re}$.

\subsection{Maturity dependence of dielectric permittivity, $\epsilon_{re}$}

From Table \ref{table 1}, the dielectric permittivity, $\epsilon_{re}$ at early stage of maturation, II, is reported as 1.9 which is lower than the data
previously reported by \citet{XJ85}; the ratios of $\epsilon_{re}$ at II to those at III, IVa and IVb, range $\sim$ 1.6 - 1.8. In II, kerogen
porosity is mainly filled by bitumen\citep{ZR15}; bitumen is mostly weakly metamorphosed at the early stage of maturation.\citep{XJ85}
For both Immature, I, and early maturation stage, II, bitumens have more normal or isomeric fatty alkanes which have high hydrogen content,i.e.,
in our case II has HI = 584, highest among all maturation stages (Table \ref{table 1}). In the stages, I and II, the molecules, atoms and electrons in 
KBC complex are , structurally, bound to each other strongly, making any distortions and orientations difficult in the presence of external electric fields. 
Therefore, the ability of polarization of 
bitumens and closely-connected molecules become extremely weak which is a potential reason for low dielectric permittivity at I and II stages.
In terms of structural defects, it is reported that weakly metamorphosed bitumen is mostly homogeneous-textured which causes interfacial and 
orientational polarization of doublet weak in strength due to lack of impurities and vacancies; this arguments support our results strongly.
In an immature stage, I, and an early maturation stage, II, insoluble (solid) kerogen component of total organic content (TOC), remains yellow-colored
(not being cooked well enough to generate oils yet). The yellow kerogen confirms any contained vitrinite particles have low reflectance,\citep{SFL82}
i.e., n$_{re}$ is low so that $\epsilon_{re}$ shows low value (as $\epsilon_{re}$ = n$_{re}^{2}$ for non magnetic materials).

$\epsilon_{re}$ display higher values ($\sim$ 1.8 times) for the maximum maturity stages in the oil-generating windows, i.e., III, IVa, IVb, 
compared to the early maturation stage, II. We observe a grouping of $\epsilon_{re}$ for III, IVa and IVb, i.e.,
$\epsilon_{re}^{(IVa)}/\epsilon_{re}^{(III)}$ $\approx$ $\epsilon_{re}^{(IVb)}/\epsilon_{re}^{(IVa)} \sim$ 1. 
\citet{ZR15} have shown that between HI = 450 to 150, basically from III through IVb,
a considerable amount of generated oil remains trapped within the shale. Possibly the variations are due to polar and non-polar
hydrocarbons still remaining in the samples. This study also indicates that an inverse correlation between HI and specific surface area (SSA)
upto HI $\approx$ 130. At higher HI ($>$ 200), SSA remains low. With solvent extraction, the case with HI $<$ 130 shows dramatic increase
in SSA. Our results point a more complicated picture of interactions among polar/non-polar free radicals, carbon contents, and HI.

In the oil window in an organic source mudrocks, the presence of considerable amount of KBC complex obtained (generated and stored).\citep{}
The state of KBC complex in III, IVa and IVb, is ready to generate liquid hydrocarbons. Kerogen starts transforming into
bitumen or other organic matters, changed coloration to brown which enhances the reflectance of any contained vitrinite particles,
i.e., $\epsilon_{re}$ increases. Both polar (aromatics, saturates, hydrogen-related groups) and non-polar (asphaltenes, nitrogen, sulfur, carbon,
oxygen-contained compounds) hydrocarbons consist in KBC complex. With increasing depth of burial, in other words, with the increasing of
maturity, nitrogen (almost all) and some of oxygen, hydrogen and carbon deplete. Further thermal cracking converts the KBC
complex into petroleum. In the oil generating windows, III, IVa and IVb, the polar hydrocarbon increases over non-polar part in the converted
hydrocarbons; bitumen is transitioned to be more carbon-condensed and aromatic as metamorphism proceeds. The aromatic-formed KBC
complex is easy to distort and orient along the applied electric field; the polarizability develops which causes $\epsilon_{re}$ risen
in the oil-expulsion main stages, III, IVa and IVb. In this high metamorphosed stages, KBC complex becomes heterogeneous (presence
of impurities and vacancies) in texture makes interfacial and orientational polarization of doublet stronger. 
The bundling of $\epsilon_{re}$ in the oil windows, can act as a demarcation line together, to illustrate
oil generating feature; if anyone goes up, one escape oil windows and move to a late or an overmatured stage, or
if anyone lies below this line, one resides in early maturation or immature stage.

In a late mature stage, V, KBC complex is highly metamorphosed and carbon-contained which means HI (= 115) goes down.
Liquid hydrocarbons (polar and non-polar) are condensed into gaseous form; bitumen is transformed into graphite (due to over carbon-formation)
and with the highest level of heterogeneity (full of impurities and vacancies).
The graphite typed-bitumen now have high electrical conductivity (of free electrons) and both interfacial and orientational polarizations (of bound 
electrons). The coloration of KBC complex turn into black and the reflectivity of any contained vitrinites must have reached highest.
Therefore, $\epsilon_{re}$ shows the highest value (= 4.9, see table \ref{table 1}). The $\epsilon_{re}$ and $\sigma_{re}$ in the
late mature stage, V, depends on only graphitization of KBC complex (due to lack of hydrocarbons)
but in oil-generating stages, III, IVa and IVb, these parameters rely on the conversions of KBC complex into polar/non-polar hydrocarbons.

\subsection{Maturity dependence of electrical conductivity, $\sigma_{re}$}

 The conductivity profiles for all stages, from II through V, are governed by $\nu^{p}$-dependence where p $\lesssim$ 1; it describes mechanisms as
 electrical polarization or dielectric relaxation. This statement is absolutley valid in our perspect because the polarizations due to bound electrons
 in the KB complex or during the interactions between charged surfaces of clay-site and KBC complex; eventhough there are few free electrons but suppressed by 
 cumulative dielectric polarizations by the large number of bound electrons and molecules. The degree of polarization determine the maturity
 stages. From Table(\ref{table 1}), we obtain $\sigma_{re}^{(V)}/\sigma_{re}^{(II)}$ = 2.8, $\sigma_{re}^{(V)}/\sigma_{re}^{(IVb)}$ = 1.4,
 and $\sigma_{re}^{(IVa)}/\sigma_{re}^{(III)}$ $\approx$ $\sigma_{re}^{(IVb)}/\sigma_{re}^{(IVa)} \sim$ 1.
 The conductivity dispersion will follow Drude-type if polar/non-polar hydrocarbons flow freely, that is definitely not the case here.
 Eventhough we have liquid hydrocarbons, petroleum, oils (from KBC complex-metamorphism) in the main oil-generating stages, III, IVa and IVb, the polar (free radicals)
 hydrocarbons interact with non-polar components, and probably charged surfaces of clay-pores, may create bindings so that they can oscillate
 and polarize along the sub-THz electrical waves we applied. However there is a chance for a few free electrons to be created and moving but it
 sould be inappreciable.

 The changes in dielectric permittivity, $\epsilon_{re}$, at different maturity stages are well-supported by their corresponding
 electrical conductivity, $\sigma_{re}$ profiles. Both $\epsilon_{re}$ and $\sigma_{re}$-data agree upon the grouping in advanced HC
 generating stages; this almost constant level emphasizes a threshold of oil-generation during which complex electrical dielectric
 constant do not change appreciably.

 The late mature or overmatured stage, V, contains condensed gas rather than free radicals (polar/non-polar hydrocarbons),
 and KBC complex is graphitized. The conductivity displays highest polarization due to free and bound electrons reside in graphites.
 The constant segment of $\sigma_{re}$ at lower frequencies, shows a smooth transition to graphite through carbon aggregation-mechanism.

 We speculate, according to Fig.(\ref{maturity_BK}), three possiblities at and beyond overmatured stage, V : if $\epsilon_{re}$ and $\sigma_{re}$ (1) increase, 
 (polar/non polar) HC are more generated, (2) stay level-off, HC are trapped, and (3) decrease, HC start releasing.

\section{Conclusions}

The most significant result we present is that the complex dielectric properties ($\epsilon_{re}$ and $\sigma_{re}$) can be strong
and reliable candidates of characterizing maturity of organic-rich mudrocks. The $\epsilon_{re}$ and $\sigma_{re}$-data emphasize that $\epsilon_{re}$ have
constant magnitudes without any dispersion but the lowest value for an early stage of maturation, II, and gradually go up with higher maturations;
$\sigma_{re}$ display sublinearilty with frequencies and lowest in magnitude at II, and increase with maturity stages (same as $\epsilon_{re}$).
Both, $\epsilon_{re}$ and $\sigma_{re}$, show the clustering in oil-windows, III, IVa and IVb, and hold the maximums for the
overmatured stage,V. In other words, dielectric permittivity and electrical conductivity support each other convincingly.
The grouping in the main HC generationg stages sketches a borderline to point out a threshold level for maximum oil expulsions, i.e., the presences
of polar/non-polar hydrocarbons and their interactions with neighbrong minerals' sites and pathways toward more carbon-formations.

Our data indicates that the kerogen-bitumen-clay (KBC) complex is one of the prime factor to be transformed during
maturation. The magnitudes of $\epsilon_{re}$ and $\sigma_{re}$ of KBC complex in the oil windows\citep{} match
exactly our data in the same maturation stages (III, IVa and IVb); these dielectric parameters stay low in magnitude at
immature or early stage of maturation and increase with maturation, supporting the patterns we obtain. The graphitization
of KBC complex at V can also be understood from the complex dielectric characterizations of organic-rich mudrocks.

\section{Acknowledgement}

This work was supported by the US Department of Energy
(Basic Energy Science) under grant DE-FG02-09ER16018. 
We also thank the OCLASSH consortium for support.

\end{document}